\documentclass[twocolumn,floatfix,showpacs,prd,aps,tightenlines,amsmath,amsfonts,amssymb]{revtex4}
\usepackage{amsmath}
\usepackage{graphicx}
\usepackage{psfrag}
\usepackage{color}
\usepackage{dcolumn}% Align table columns on decimal point
\usepackage{bm}% bold math
\usepackage{longtable}
\usepackage[mathscr]{eucal}
\usepackage{mathrsfs}

\newcommand{\bea}{\begin{eqnarray}}
\newcommand{\eea}{\end{eqnarray}}
\newcommand{\beq}{\begin{equation}}
\newcommand{\eeq}{\end{equation}}

\begin{document}

\def\fun#1#2{\lower3.6pt\vbox{\baselineskip0pt\lineskip.9pt
  \ialign{$\mathsurround=0pt#1\hfil##\hfil$\crcr#2\crcr\sim\crcr}}}
\def\lap{\mathrel{\mathpalette\fun <}}
\def\gap{\mathrel{\mathpalette\fun >}}
\def\kms{{\rm km\ s}^{-1}}
\def\vk{V_{\rm recoil}}

\title{Study of Conformally Flat Initial Data for Highly Spinning Black Holes\\
and their Early Evolutions}

\author{
Carlos O. Lousto,
Hiroyuki Nakano,
Yosef Zlochower,
Bruno C. Mundim,
Manuela Campanelli
}

\affiliation{Center for Computational Relativity and Gravitation,\\
and School of Mathematical Sciences, Rochester Institute of
Technology, 85 Lomb Memorial Drive, Rochester, New York 14623}

\begin{abstract}
We study conformally-flat initial data for an arbitrary number of
spinning black holes with exact analytic solutions to the momentum
constraints constructed from a linear combination of the classical
Bowen-York and conformal Kerr extrinsic curvatures.  The solution
leading to the largest intrinsic spin, relative to the ADM mass of the
spacetime $\epsilon_S=S/M^2_{\rm ADM}$, is a superposition with
relative weights of $\Lambda=0.783$ for conformal Kerr and $(1-\Lambda)=0.217$
for Bowen-York.  In addition, we  measure the spin relative to the
initial horizon mass $M_{H_0}$, and find that the quantity
$\chi=S/M_{H_0}^2$ reaches
a maximum of $\chi^{\rm max}=0.9856$ for $\Lambda=0.753$.  After
equilibration, the final black-hole spin should lie in the interval
$0.9324<\chi_{\rm final}<0.9856$.  We perform full numerical
evolutions to compute the energy radiated and the final horizon mass
and spin.  We find that the black hole settles to a final spin of
$\chi_{\rm final}^{\rm max}=0.935$ when $\Lambda=0.783$. We also study
the evolution of the apparent horizon structure of this {\it maximal}
black hole in detail.
\end{abstract}

\pacs{04.25.dg, 04.30.Db, 04.25.Nx, 04.70.Bw}
\maketitle

\section{Introduction}

With the breakthroughs in the numerical techniques to evolve
black-hole binaries (BHBs)~\cite{Pretorius:2005gq, Campanelli:2005dd,
Baker:2005vv}, Numerical Relativity (NR) has become a very important tool
to explore highly-dynamical and nonlinear predictions of General
Relativity.  NR can now be used to evolve generic black-hole binaries
and accurately compute gravitational radiation from such
systems. However, these computations  are very costly when one
explores various corners of the BHB parameter space, such as extreme
mass ratios and nearly maximally spinning black holes. Yet those cases
are of great astrophysical interest~\cite{Gou:2011nq}.  
Recently, BHBs with a mass ratio
of 1:100 were successfully simulated~\cite{Lousto:2010ut,
Nakano:2011pb, Sperhake:2011ik}, and equal-mass BHBs with spin
parameters up to at least $0.97$ of the maximum spin were evolved for 25
orbits~\cite{Lovelace:2011nu}.  These latter simulations used the
first order generalized harmonic formalism and excision of the
black-hole (BH) interiors in a pseudospectral evolution scheme.

Simulations of highly-spinning black holes are important for
understanding astrophysical black holes.
Important spin effects have been found in the merger of BHBs,
such as the hangup effect \cite{Campanelli:2006uy}, 
large recoil velocities of up to 5000km/s \cite{Lousto:2011kp}, 
and their statistical consequences for galaxy merger \cite{Lousto:2012su}. 
Recent observational evidence for highly recoiling black holes 
was reviewed in Ref. \cite{Komossa:2012cy}.

Many of the groups in numerical relativity have adopted 
the so called, moving punctures
formalism~\cite{Campanelli:2005dd, Baker:2005vv}, where the interiors
of the black holes are evolved without imposing internal boundary conditions.
Since this approach usually starts from conformally flat
puncture~\cite{Brandt97b} initial conditions, and since there is a proof 
(under the assumption of axisymmetry and that such a slicing should
vary smoothly with the spin parameter $a$) of the
nonexistence of conformally flat slices of a Kerr black
hole~\cite{Garat:2000pn}, it is generally believed that these data cannot
represent highly-spinning black holes.  However, the above statement is not
quantitative, nor does it take into account the other ``half'' of the
initial data,  i.e. the extrinsic curvature. Even if we could find 
slices that make the Kerr metric conformally flat, they would
presumably not be
maximal, i.e. $K\not=0$. This means that the Hamiltonian and 
momentum constraints would couple, making the latter nonlinear  and
superposition of solutions no longer valid.

The first and most widely used initial extrinsic curvature data for spinning
black holes is the Bowen-York (BY) solution~\cite{Bowen80} to the
momentum constraints. This assumes a conformally flat 3-metric and
a longitudinal, trace-free extrinsic curvature. 
Cook and York~\cite{Cook:1989fb}
studied intrinsic spins of these data [normalized to the square of 
Arnowitt-Deser-Misner (ADM) mass~\cite{Arnowitt62} (ADM)
mass of the spacetime], $\epsilon_S=S/M_{\rm ADM}^2$, and found
that $\epsilon_S$ approaches a maximum value of $\epsilon^{\rm
max}_{S\, \rm BY}=0.928$. Further increases of the spin for these data
leads to a corresponding increase in the ADM mass such that the
intrinsic spin is unchanged.
Later, in Ref.~\cite{Dain:2002ee}, it was shown
that the conformal extrinsic curvature of a Kerr BH was also an exact
solution to the conformally flat momentum constraints. We denote such
initial data by ``cKerr'' to distinguish them from the true Kerr initial
data. Those solutions
have a maximum intrinsic spin of $\epsilon^{\rm max}_{S\, \rm cKerr}=0.932$.
This raises the possibility that, by choosing a different extrinsic
curvature, one can reach still higher maximum values of the intrinsic
spin, while keeping the conformal flatness condition for the initial
3-metric.  In this paper we will explore this possibility explicitly.
Assuming the optimal extrinsic curvature is still close to
that of a Kerr BH, we will study a parametrization of deviations
from the conformal Kerr extrinsic curvature proportional to the 
difference between the conformal Kerr and BY extrinsic curvatures.
This plays the role of a sourceless extrinsic curvature
{\it wave}.  We also perform high resolution evolutions that allow us to
shed some light on
the transition from the near maximally spinning conformally flat initial 
data to the final submaximal spinning Kerr black hole. In particular we
track the evolution of trapped surfaces and apparent horizons. 

\section{Initial Value Problem for Black Holes}

We assume a conformally flat 3-metric for the initial configuration
\bea
\gamma_{ij} &=& \psi^4 \, \tilde{\gamma}_{ij} \,,
\eea
where $\tilde{\gamma}_{ij}$ is a conformal 3-metric
and the 3-Ricci tensor $\tilde{R}_{ij}=0$.
The physical extrinsic curvature is given by 
$K_{ij} = \psi^{-2} \tilde{K}_{ij}$, and $K^{ij} = \psi^{-10} \tilde{K}^{ij}$.
The indices are raised and lowered with the conformal 3-metric $\tilde{\gamma}_{ij}$.
$\tilde{\nabla}_i$ is the covariant derivative
with respect to the conformal 3-metric $\tilde{\gamma}_{ij}$.

Assuming the maximal slicing $\tilde{K}=\tilde{K}_{i}{}^{i}=0$,
the momentum constraint becomes,
\bea
\tilde{\nabla}^j \tilde{K}_{ij} &=& 8\pi \tilde{j}_i \,,
\label{eq:mom_con}
\eea
where $\tilde{j}_i$ is a conformal matter momentum density,
and the physical matter momentum density is given
by $j_i = \psi^{-6} \tilde{j}_i$.

The Hamiltonian constraint the leads to
\bea
  \Delta \psi + \frac{1}{8}\tilde K_{ij} \tilde K^{ij} \psi^{-7} = 0 \,,
\eea
where, for conformally flat initial data, $\Delta$ is the ordinary
Laplacian operator. Taking $\psi=1 + m_p/(2 r) + u(\vec x)$, where
$m_p$ is a parameter, $r=|\vec x|$, and $u(\vec x)$ is continuous function (in fact,
$C^2$)
leads to unique solutions of the Hamiltonian constraint.

To solve the momentum constraint in Eq.~(\ref{eq:mom_con}), we split the
conformal extrinsic curvature $\tilde K_{ij}$ into a
transverse-traceless and longitudinal part~\cite{Cook:2000vr}
of the form
\bea
\tilde{K}_{ij} &=& (\mathbb{L} V)_{ij} + T_{ij} \,,
\label{eq:VT}
\eea
where the operator $\mathbb{L}$ is given by 
\bea
(\mathbb{L} V)_{ij} &=& 2 \, \tilde{\nabla}_{(i} \, V_{j)} 
- \frac{2}{3} \, \delta_{ij} \, \tilde{\nabla}_k \, V^k 
\,,
\eea
and $T_{ij}$ is a symmetric, transverse-traceless tensor,
\bea 
\tilde{\nabla}^j \, T_{ij} &=& 0 \,,
\quad
T_{i}{}^{i} = 0 \,.
\eea
Here  $V_i$ must be a solution of
\bea
\tilde{\nabla}^j  (\mathbb{L} V)_{ij} &=&  8\pi \tilde{j}_i \,.
\eea

The spinning BY extrinsic curvature~\cite{Bowen80}
is obtained from the vector field
\bea
V^i &=& \frac{1}{r^2} \epsilon^{ijk} n_j S_k \,,
\eea
where $n^i=x^i/r$ in the Cartesian coordinates $\{ x,\,y,\,z \}$, 
$S^i$ is a spin vector, 
and the Levi-Civita symbol is defined as $\epsilon^{xyz}=\epsilon_{xyz}=1$.
When we include $r=0$ in this analysis, 
the momentum constraint becomes
\bea
\tilde{\nabla}^j \tilde{K}^{\rm BY}_{ij} &=& 4\pi \epsilon_{ijk} S^k \tilde{\nabla}^j 
\delta^{(3)}( x^{\ell} ) \,.
\label{eq:sourceBY}
\eea
If we do not add an additional $T_{ij}$ part, the spinning BY solution is a purely longitudinal solution.

To solve for these initial data numerically, we use the {\sc
TwoPunctures} thorn~\cite{Ansorg:2004ds} and the
{\sc Cactus}/{\sc Carpet}/{\sc EinsteinToolkit} infrastructure~\cite{cactus_web,
einsteintoolkit, Loffler:2011ay, Schnetter-etal-03b}.
We restrict our analysis here to single BH spacetimes. We solve the
Hamiltonian constraint for $\psi$ using the {\sc TwoPunctures} thorn by
setting the mass and spin of the second puncture to zero.

We use {\sc AHFinderDirect}~\cite{Thornburg2003:AH-finding} to locate
apparent horizons (AHs).  We measure the magnitude of the horizon spin using
the Isolated Horizon algorithm detailed in Ref.~\cite{Dreyer02a}.
Note that once we have the
horizon spin, we can calculate the horizon mass via the Christodoulou
formula
\begin{equation}
{m_H} = \sqrt{m_{\rm irr}^2 + S_H^2/(4 m_{\rm irr}^2)} \,,
\end{equation}
where $m_{\rm irr} = \sqrt{A/(16 \pi)}$ and $A$ is the surface area of
the horizon, and $S_H$ is the spin angular momentum of the BH (in
units of $M^2$). We denote the horizon mass on the
initial slice and the final equilibrated horizon mass by
$M_{H_0}$ and $M_{H_\infty}$, respectively.

\section{Generalization of Conformally-Flat Initial-Data}

In Ref.~\cite{Dain:2002ee} it was shown that
\bea
\tilde{K}_{ij} &=& 
\frac{2}{r^2 \sin^4 \theta} \bigl[ 
(\hat{S}^a \tilde{\nabla}_a \omega) \epsilon_{k \ell (i} \, \hat{S}^k \,n^\ell \,n_{j)}
\nonumber \\ && \qquad
-
(n^a \tilde{\nabla}_a \omega) \epsilon_{k \ell (i} \, \hat{S}^k \,n^\ell \,\hat{S}_{j)}
\bigr]
\label{eq:Komega}
\eea
is
a solution to the momentum constraints,
where $\cos \theta = \hat{S}^i n_i$ and  $\hat{S}^i = S^i/S$, provided
that the coordinates are (quasi-) isotropic (i.e.\ the conformal
metric is flat, or conformally Kerr in quasi-isotropic coordinates).
The (otherwise arbitrary) scalar function $\omega$ gives the angular 
momentum of the BH~\cite{Bowen80}. That is, the spin angular momentum
$J_i$ is given by
\bea
J_i  &=& \frac{1}{16\pi} \epsilon_{ijk} \oint_{r=\infty}
(x^j K^{km} - x^k K^{jm}) \, d^2 S_m
\nonumber \\ 
&=& - \frac{\hat{S}_i}{4} \Big( \omega(\theta=\pi) 
- \omega(\theta=0) \Big) \Big|_{r=\infty} \,,
\label{eq:Ji}
\eea
assuming  $\psi \to 1$ in the limit $r \to \infty$.

The spinning BY initial data can be recovered by taking
\bea
\omega_{\rm BY} = - S \left( \cos^3 \theta - 3\,\cos \theta \right) \,.
\eea
On the other hand, Eq.~(\ref{eq:Komega}) is linear in $\omega$, and we
can therefore take an arbitrary superposition $\sum_i c_i \,\omega_i$,
where $c_i$ are constants and each $\omega_i$ yields a solution to the
vacuum momentum constraint. In particular, we are interested in
\bea
\delta \omega = \omega^{\rm K} - \omega^{\rm BY}
= \frac{M_{\rm K} \,a^3 \,\sin^4 \theta \,\cos \theta}{\Sigma} \,,
\eea
where $a=S_{\rm wave}/M_{\rm K}$, 
$\Sigma=({r_{\rm BL}}^2+a^2\,\cos^2\theta)$
and 
$r_{\rm BL} =  r [1+(M_{\rm K}+a)/(2r)] [1+(M_{\rm K}-a)/(2r)]$.
Here we denote the spin parameter by $S_{\rm wave}$ rather than $S$
because it is in a transverse-traceless ``wave-like'' part 
of the extrinsic curvature (see below), and does
not contribute to the ADM angular momentum (unlike $S$).
Therefore, it need not match the BH's spin.

Replacing 
$\omega$ in Eq.~(\ref{eq:Komega}) by $\delta \omega$ yields a solution
to the momentum constraints with zero spin, which we will denote by
$\delta \tilde K_{ij}=\tilde K_{ij}^{\rm K} - \tilde K_{ij}^{\rm BY}$. 
We note that $\delta \tilde K_{ij}$
is transverse-traceless, 
i.e., $\tilde{\nabla}^j \delta \tilde{K}_{ij} = 0$ everywhere,
because it is constructed from the difference of two extrinsic
curvatures each leading to the same source term $\tilde j_i$ in 
Eq.~(\ref{eq:mom_con}).
Finally, the extrinsic curvature given 
by $\tilde K_{ij} = \tilde K_{ij}^{\rm BY} + \Lambda \delta \tilde
K_{ij}$ solves the momentum constraints, with non-zero spin, for any
fixed value of $\Lambda$. 
Taking $\Lambda=1$ yields a pure conformal
Kerr extrinsic curvature (provided that one uses the same spin
parameter $S_{\rm wave} = S$ to construct  $\tilde K_{ij}^{\rm BY}$  and $\delta \tilde
K_{ij}$), while $\Lambda=0$ corresponds to pure BY
curvature. Again, we note that $\delta \tilde
K_{ij}$ does not contribute to the spin, so in principle, one can use
a different spin parameter to construct  $\tilde K_{ij}^{\rm BY}$ and
$\delta \tilde
K_{ij}$.

\section{Analysis of single spinning black hole initial data}

The extrinsic curvature in Eq.~(\ref{eq:Komega}) can be written as
\bea
\tilde{K}_{ij} &=& 
\frac{2}{r} \bigl[ 
\bigl( \alpha_{\theta} - (\hat{S}^m n_m)^2 \alpha_r \bigr)
\epsilon_{k \ell (i} \, \hat{S}^k \,n^\ell \,n_{j)}
\nonumber \\ && \qquad
+ \alpha_r (\hat{S}^m n_m)
\epsilon_{k \ell (i} \, \hat{S}^k \,n^\ell \,\hat{S}_{j)}
\bigr]
\,,
\label{eq:Kaa}
\eea
where the coefficients $\alpha_{\theta}$ and $\alpha_r$ are given by
\bea
\alpha_{\theta}^{\rm BY} &=& \frac{3S}{r^2} \,,
\quad \alpha_{r}^{\rm BY} = 0 \,,
\label{eq:alpha_BY}
\eea
for the BY case, and for $\delta \omega$ we have
\bea
\delta \alpha_{\theta} &=&
-16\,M_{\rm K}\,a^3
 \bigl[ 
 ( -40\,r M_{\rm K}{a}^{2}+5\,{a}^{4}+5\,{M_{\rm K}}^{4}
\nonumber \\ && \quad
-10\,{M_{\rm K}}^{2}{a}^{2}+80\,{r}^{4}
-24\,{r}^{2}{a}^{2}+40\,r{M_{\rm K}}^{3}
\nonumber \\ && \quad
+160\,{r}^{3}M_{\rm K}+120\,{r}^{2}{M_{\rm K}}^{2} ) \,\cos^2 \theta
\nonumber \\ && \quad
- \left( 2\,r+M_{\rm K}+a \right) ^{2} \left( 2\,r+M_{\rm K}-a \right) ^{2}
\nonumber \\ && \quad
+ 48\, {a}^{2}{r}^{2}\,\cos^4 \theta 
 \bigr] / \beta^2
\,,
\nonumber \\ 
\delta \alpha_r &=& 
32\,M_{\rm K}\,a^3
 \left( 2\,r+M_{\rm K}+a \right)  \left( 2\,r+M_{\rm K}-a \right)  
\nonumber \\ && \quad \times
\left( {a}^{2}-{M_{\rm K}}^{2}+4\,{r}^{2} \right)
/ \beta^2
\,,
\label{eq:delta_alpha}
\eea
where
\bea
\beta &=& 16\,r^2\,\Sigma
\nonumber \\ 
&=& 
16\,{r}^{4}+{M_{\rm K}}^{4}+{a}^{4}+24\,{r}^{2}{M_{\rm K}}^{2}
-8\,{r}^{2}{a}^{2}
\nonumber \\ &&
+8\,r{M_{\rm K}}^{3}-2\,{M_{\rm K}}^{2}{a}^{2}
-8\,rM_{\rm K}{a}^{2}
\nonumber \\ &&
+32\,{r}^{3}M_{\rm K}
+16\,{a}^{2}{r}^{2}\,\cos^2 \theta  \,.
\eea

As a way to study the effects of changing the extrinsic curvature on
the maximum intrinsic spins, we 
set the conformal extrinsic curvature to
\bea
\tilde{K}_{ij} = \tilde{K}_{ij}^{\rm BY} +
\Lambda\, \delta \tilde{K}_{ij} \,,
\label{eq:lambda}
\eea
and vary the parameters
$\Lambda$, $M_{\rm K}$, and the spin parameter $S_{\rm wave}=M_{\rm K} a$
used to construct $\delta \tilde{K}_{ij}=(\tilde K_{ij}^{\rm K} - \tilde K_{ij}^{\rm BY})$.
In all cases the spin used to construct
$\tilde{K}_{ij}^{\rm BY}$ remains fixed at $S=M^2$.

\subsection{ADM mass dependence}

\begin{figure}[!ht]
\includegraphics[width=\columnwidth]{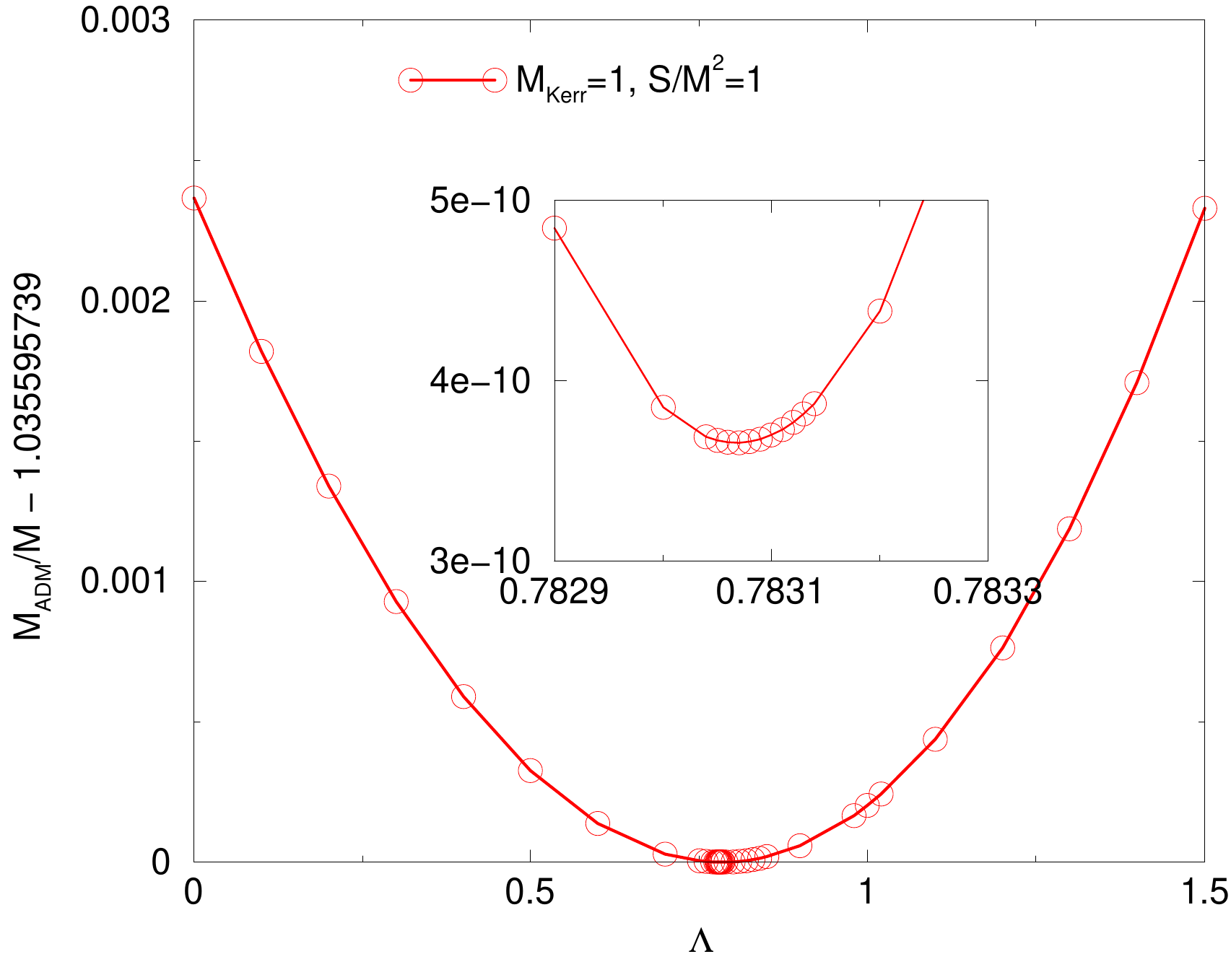}
\caption{The ADM mass of a single spinning BH with spin parameter
$S=M^2$ versus weight $\Lambda$. Here the spin of the wave $\delta
\tilde{K}_{ij}$ is fixed at $S_{\rm wave}=M^2$. Note that $\Lambda=1$, which
corresponds to conformal Kerr extrinsic curvature, does not lead to the minimum
mass, and hence maximum intrinsic spin.}
\label{fig:aseq}
\end{figure}

\begin{figure}[!ht]
\includegraphics[width=\columnwidth]{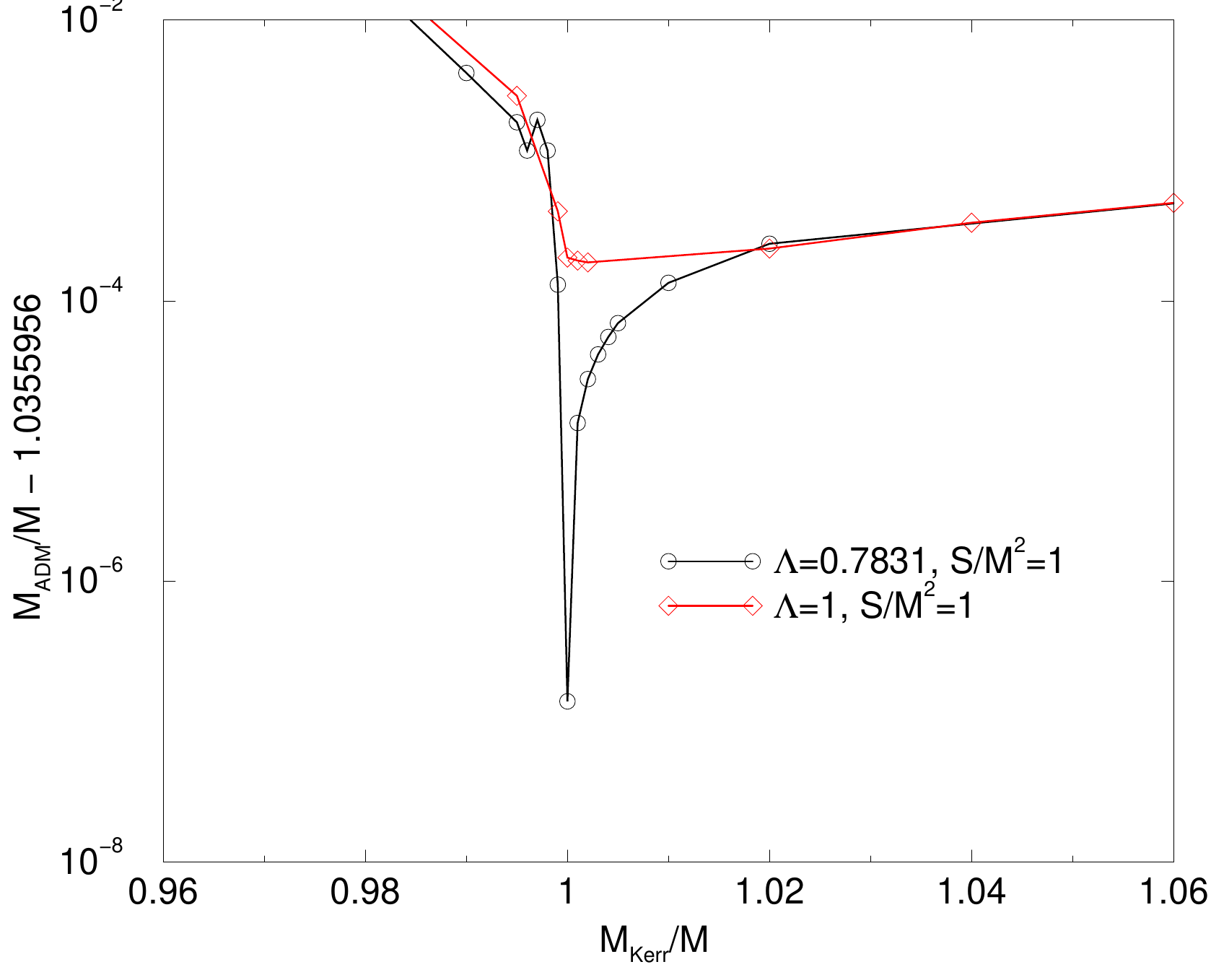}
\caption{The ADM mass of a single spinning BH with spin parameter
$S=M^2$ as a function of the Kerr mass parameter in the
extrinsic curvature wave. Here the spin of the wave $\delta
\tilde{K}_{ij}$ is fixed at $S_{\rm wave}=M^2$. A minimum in the ADM mass is obtained
for $M_{\rm K}=1$
when $\Lambda=0.7831$. The minimum is at slightly larger values of
$M_{\rm K}$ for pure conformal Kerr extrinsic curvature.}
\label{fig:mseq}
\end{figure}

\begin{figure}[!ht]
\includegraphics[width=\columnwidth]{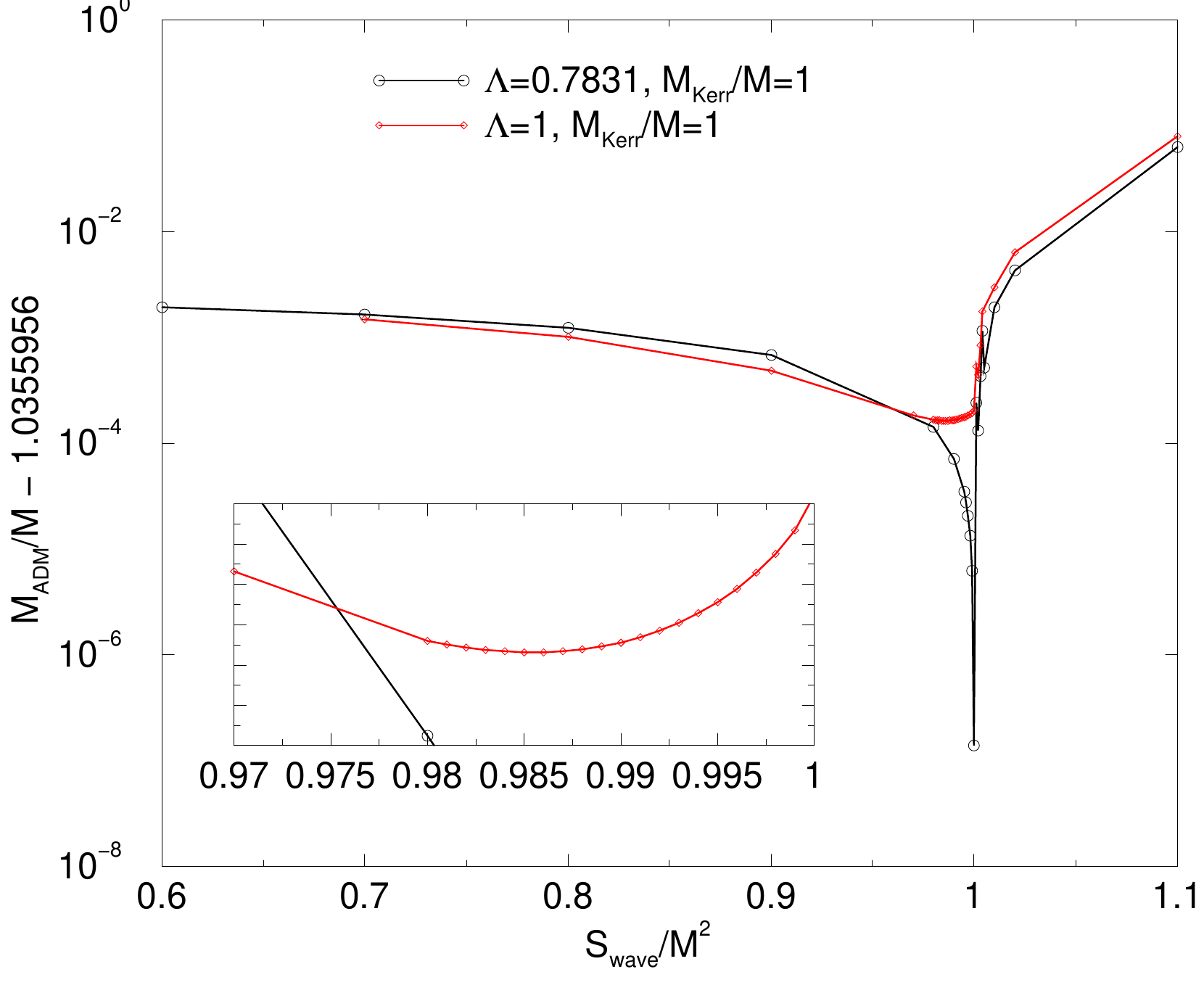}
\caption{The ADM mass of a spinning BH with spin parameter
$S=M^2$ as a function of spin of the curvature wave $S_{\rm wave}$
for $\Lambda=0.7831$ and $\Lambda=1$.
Here $M_{\rm K}=M$.}
\label{fig:sseq}
\end{figure}

To estimate the final (equilibrated) the intrinsic spin for a large
number of configurations, we measure the ADM mass of a single
spinning BH spacetime with spin parameter $S=M^2$ and puncture
 mass $m_p=0.02M$.
We choose this puncture mass so that it has a negligible effect on the
total ADM mass, while still leading to a solvable initial data problem
(taking $m_p=0$ would lead to singularities at the puncture). For
the sequences presented here, we only perform numerical evolutions
to calculate the equilibrated spin on a select set of configurations. 
For the rest, we note that equilibrated value of the intrinsic spin
$S_H/M_{H_\infty}^2$ (as measured using the equilibrated horizon mass) must
be larger than $S_H/M_{\rm ADM}^2$, which holds true because these
axisymmetric data can radiate mass but not angular momentum. Since
the total mass radiated is small
 ($\delta M / M <0.2\%$)~\cite{Dain:2002ee}, $S_H/M_{\rm
ADM}^2$ is a reasonable approximation for the final spin. One might expect
that the conformal Kerr extrinsic curvature leads to the largest intrinsic spin,
but as shown in Fig.~\ref{fig:aseq}, the actual minimum in the ADM
mass occurs at $\Lambda=0.7831$, or roughly a mixture of 78\% conformal Kerr
and 22\% BY extrinsic curvatures. As seen in Fig.~\ref{fig:mseq}, if we take $\Lambda=0.7831$ and vary the
mass parameter $M_{\rm K}$ in the extrinsic curvature (note that this is not
the mass of the BH, rather it is the proportionality factor between
$S_{\rm wave}$ and the Kerr $a$ parameter in the extrinsic
curvature $\delta \tilde{K}_{ij}$) 
we find that the ADM mass is minimized when $M_{\rm K}=M$. On the
other hand, if we take $\Lambda=1$, the ADM mass is minimized at
slightly larger values of $M_{\rm K}$. Finally, we need not have the spin of
the wave (i.e.\ the spin parameter in $\delta \tilde{K}_{ij}$) match the BY
spin. In Fig.~\ref{fig:sseq} we plot the ADM mass for a BH of spin
$S=M^2$ and vary the ``spin'' of the curvature wave. The minimum in
the ADM mass occurs when $S_{\rm wave}=S$ when $\Lambda=0.7831$ and at
slightly smaller values for $\Lambda=1$ (note that $\Lambda=1$ only
correspond to pure conformal Kerr data when $S_{\rm wave}=S$).

The smallest ADM mass, and therefore the largest intrinsic spins, 
are obtained when $S_{\rm wave}=S$, $M_{\rm K}=M$, and $\Lambda=0.7831$. The
corresponding spin is $S_H/M_{\rm ADM}^2 = 0.9324$, which is
slightly larger than the maximum spin of $0.932$ obtained from purely
conformal Kerr extrinsic curvature.


\subsection{Horizon mass dependence}

\begin{figure}[!ht]
\includegraphics[width=\columnwidth]{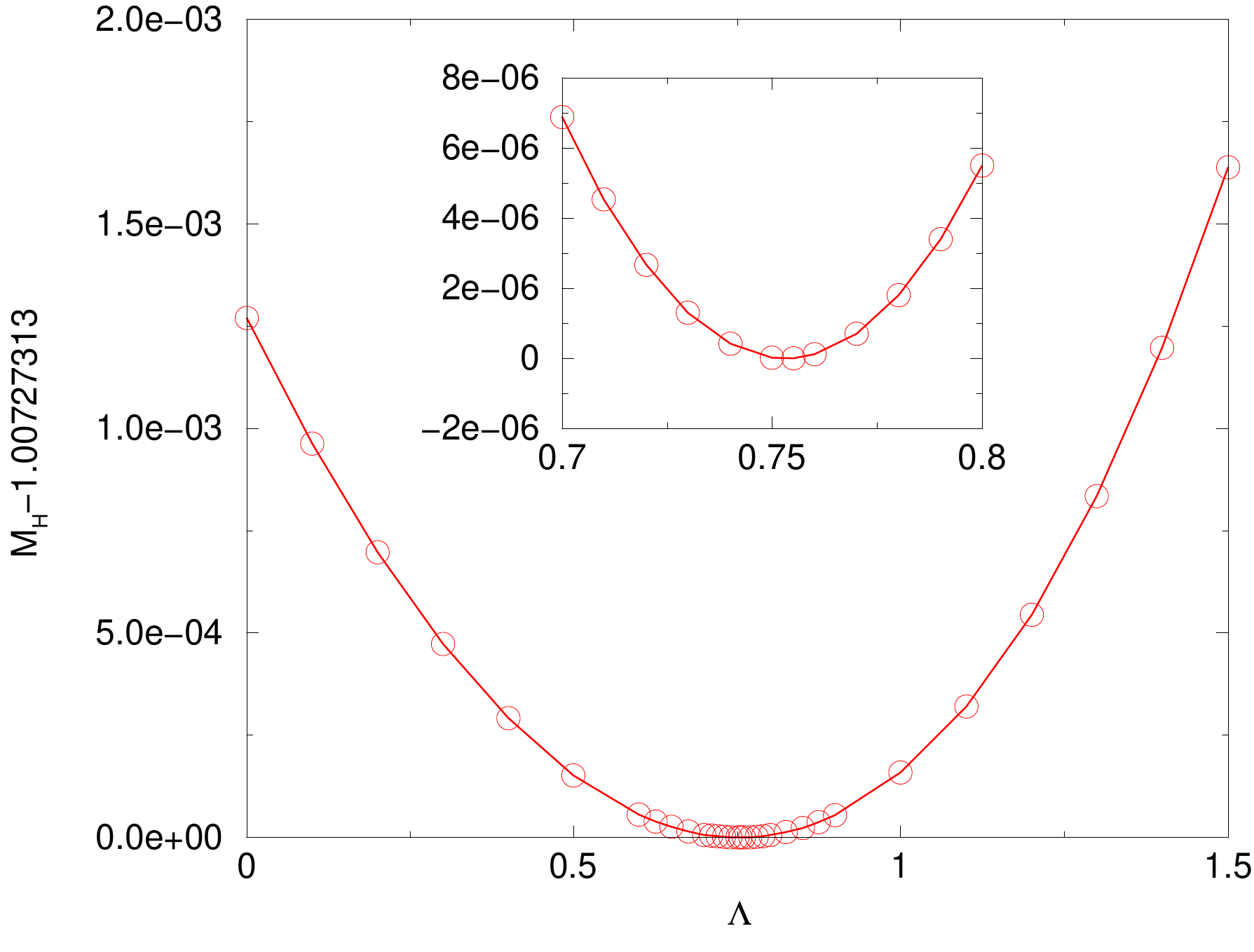}
\caption{The horizon mass of a spinning BH with spin parameter
$S=M^2$ as a function of $\Lambda$, the mixing parameter for BY and conformal Kerr
curvatures.}
\label{fig:mh}
\end{figure}

The difference between the ADM mass and the initial horizon mass of
the 
black holes
provides a measure of the energy initially lying outside the horizon. This
energy can potentially escape to infinity or be absorbed by the black hole.
We thus expect that the evolution of these  spinning black holes 
will lead to a stationary black hole with 
\bea
\epsilon_{S\,{\rm initial}}=\frac{S}{M_{\rm ADM}^2}<\chi_{\rm
final}<\chi_{\rm initial}=\frac{S}{M_{H_0}^2}
\,.
\eea
For the $\Lambda^{\rm max}=0.7831$ case,
$0.9324<\chi_{\rm final}<0.9856$. We also investigate the minimum of
the horizon mass when $\Lambda$ is varied but $S$ and $M_k$ are set
$M^2$ and $M$, respectively. The results are
shown in Fig.~\ref{fig:mh}. The minimum horizon mass  occurs
when $\Lambda^{\rm max}=0.753$,
which is close to where the minimum in the ADM mass occurs.  We
summarize the initial parameters in Table~\ref{tab:IDparameters}.

\begin{table}[t]
  \caption{Initial spin parameters for Bowen-York data (BY),  the new data at
 $\Lambda^{\rm max}$, and conformal Kerr data. Note that $\xi=\chi/(1+\sqrt{1-\chi^2}$). In all cases, the spin $S/M^2$ is identically 1.}
   \label{tab:IDparameters}
\begin{ruledtabular}
\begin{tabular}{|l|l|l|l|}
ID	& $\epsilon_S=S/M^2_{\rm ADM}$ & $\chi=S/M_{H_0}^2$ & $\xi=S/(2M^2_{\rm irr})$ \\
\hline
BY 	&  0.9282		   & 0.9831   & 0.8311	 \\
$\Lambda^{\rm max}$ & 0.9324	   & 0.9856   & 0.8431\\
cKerr & 0.93207  & 0.9854 & 0.8421\\
\end{tabular}
\end{ruledtabular}
\end{table}

\section{Evolutions}

We evolve these single (distorted) 
BH data-sets using the {\sc
LazEv}~\cite{Zlochower:2005bj} implementation of the moving puncture
approach~\cite{Campanelli:2005dd,Baker:2005vv} with the conformal
function $W=\exp(-2\phi)$ suggested by Ref.~\cite{Marronetti:2007wz}.
For the runs presented here,
we use centered, eighth-order finite differencing in
space~\cite{Lousto:2007rj} and a fourth-order Runge Kutta time integrator. (Note that we do
not upwind the advection terms.)

Our code uses the {\sc Cactus}/{\sc EinsteinToolkit}~\cite{cactus_web,
einsteintoolkit} infrastructure.
For these runs, we used the {\sc Carpet}~\cite{Schnetter-etal-03b} mesh refinement driver to
provide fixed mesh refinement (since the BHs do not move across the
grid); initially starting with 15 levels of refinement.
 When a larger AH forms, we reduce the numbers of levels from 
15 to 11. Our base (coarsest) grid extended from $-400M$ to $400M$ in
all directions with a resolution of $h=10M/3$. We used the $\pi$
 and $z-$reflection
symmetries to reduce the computational domain by a factor of 4. The
resolution on the  finest grid was $h=M/1000$. We took a constant
Courant factor of $dt/h=1/4$ in all grids. We were able to achieve
good spin preservation by choosing a grid structure where the finest
grid is roughly twice as wide as the AH.
Here
we extended to finest grid to $\pm0.021M$.

We found that using a lower-order dissipation operator, in this case
fifth-order, gave sufficient accuracy, while, at the same time,
reducing the number of required buffer points at the refinement
boundary, as well as the cost in walltime of the dissipation step. At
these boundaries, we used 16 buffer points. Standard ninth-order 
dissipation would require 20 buffer zones, while seventh-order would
not require additional buffer zones,  it proved to give a more noisy
waveform than using the fifth-order dissipation stencil.

The initial data are axisymmetric, which we exploit in order to reduce
the number of spectral collocation points to $140\times140\times4$.
We then use the full spectral expansion, rather than the much faster
interpolation techniques, to transfer the initial data to the
numerical grid.

We obtain accurate, convergent waveforms and horizon parameters by
evolving this system in conjunction with a modified 1+log lapse and a
modified Gamma-driver shift
condition~\cite{Alcubierre02a,Campanelli:2005dd,vanMeter:2006vi}, and an initial lapse
$\alpha(t=0) = 2/(1+\psi_{BL}^{4})$.  The lapse and shift are evolved
with
\begin{subequations}
\label{eq:gauge}
  \begin{eqnarray}
(\partial_t - \beta^i \partial_i) \alpha &=& - 2 \alpha K,\\
 \partial_t \beta^a &=& (3/4) \tilde \Gamma^a - \eta \beta^a \,,
 \label{eq:Bdot}
 \end{eqnarray}
 \end{subequations}
where we use $\eta=1$ for all simulations presented below.

\begin{figure}[!ht]
  \includegraphics[width=\columnwidth]{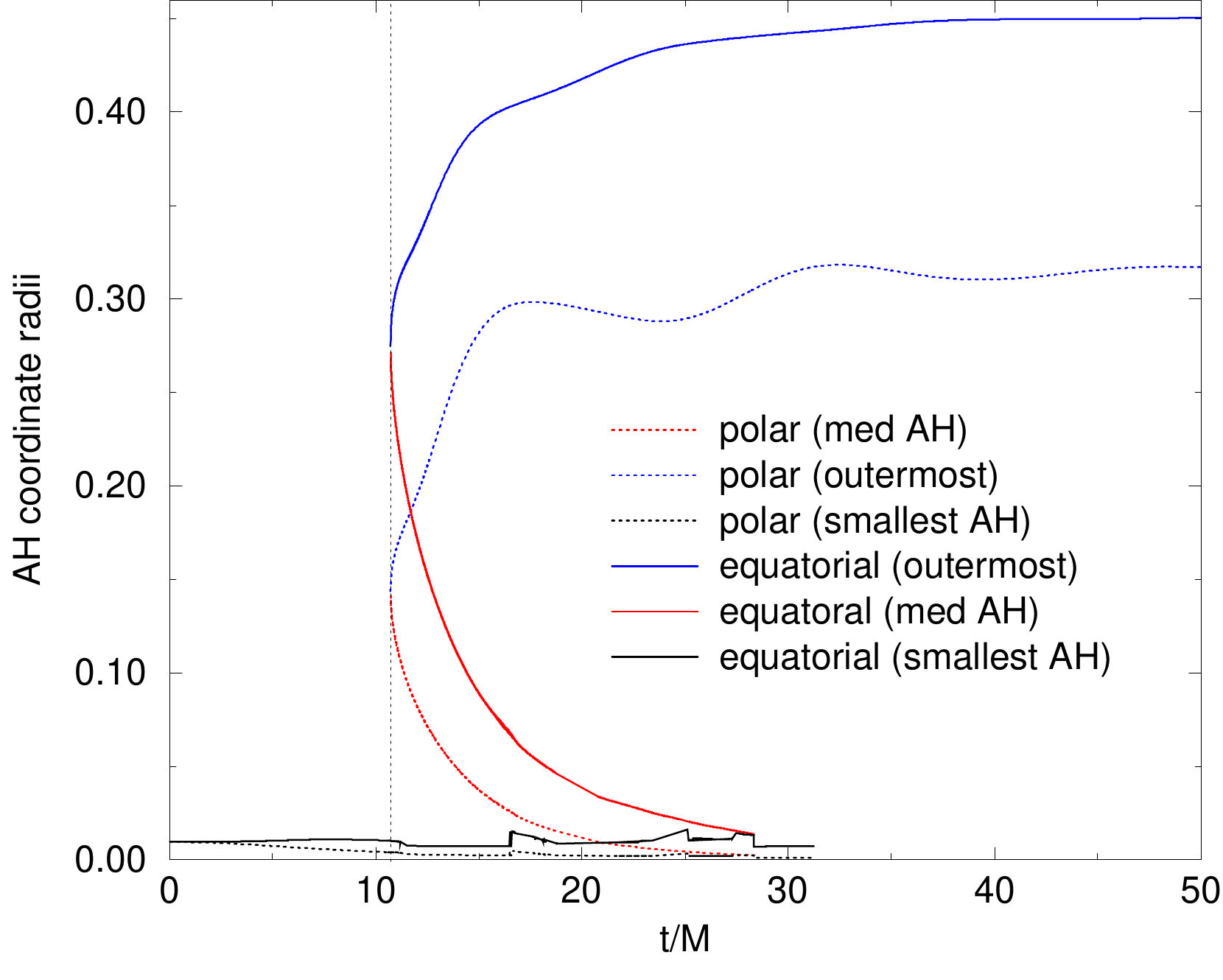}
  \includegraphics[width=\columnwidth]{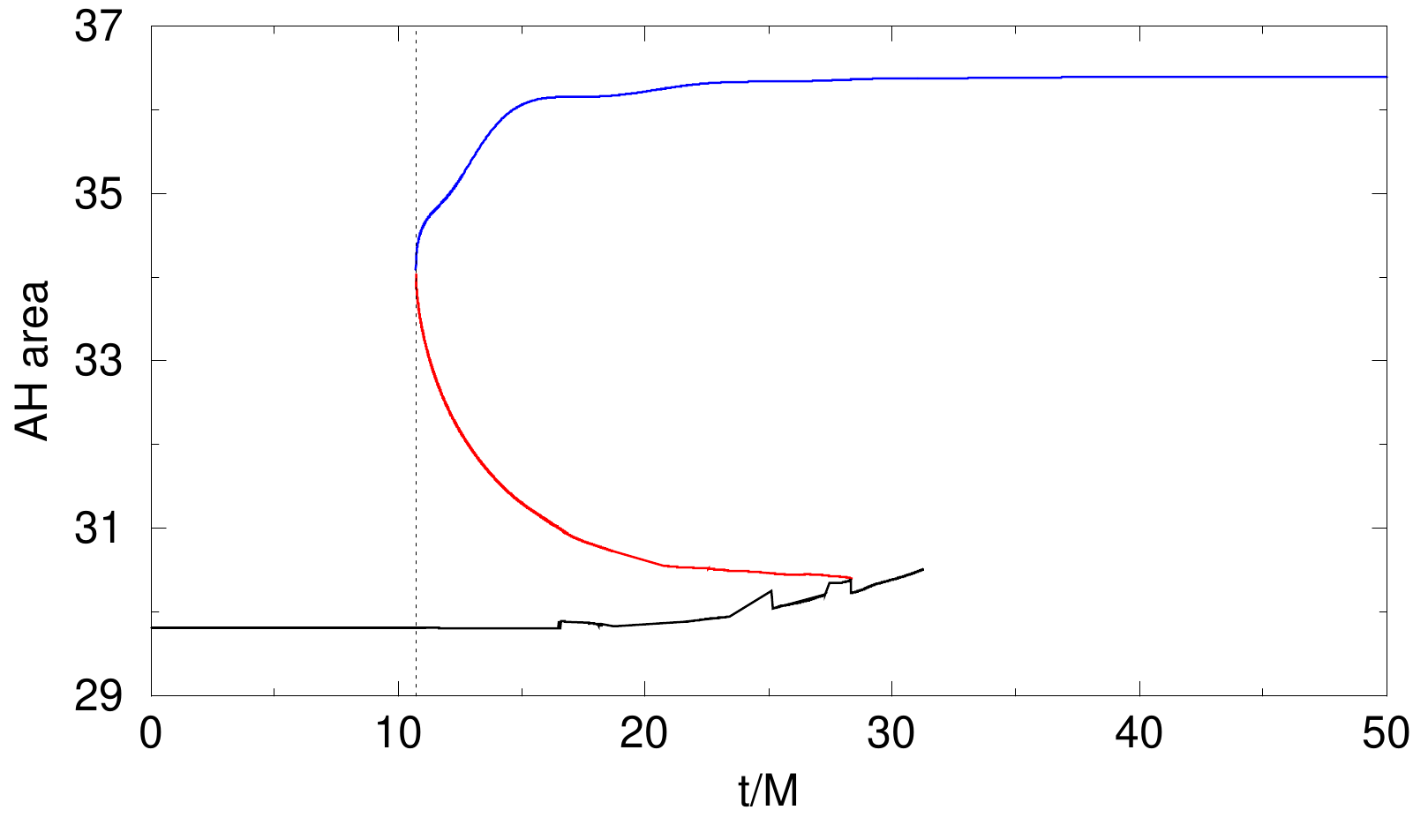}
  \caption{(Top) The polar and equatorial radii of the three AHs
  found  for the $\Lambda=0.7831$ data. (Bottom) The area of the same 
three AHs. All in units of the ADM mass. The dotted vertical
line at $t=10.7M$ represents the time when an external
trapped surface first forms.
}
\label{fig:AH}
\end{figure}

The initial AH spherical and very small (see Fig.~\ref{fig:AH}),
with coordinate radius $r\sim0.009$. A larger AH forms after the BH
absorbs the excess radiation on the grid, leading to an oblate
spheroid AH with
coordinate radius $\sim0.3-0.5$. We drop levels of refinement once
this larger AH begins to equilibrate.

We measure radiated energy in
terms of the radiative Weyl scalar $\psi_4$, using the formulas provided in
Refs.~\cite{Campanelli:1998jv,Lousto:2007mh}. However, rather than using
the full $\psi_4$, we decompose it into $\ell$ and $m$ modes and solve
for the radiated linear momentum, dropping terms with $\ell \geq 5$.
The formulas in Refs.~\cite{Campanelli:1998jv,Lousto:2007mh} are valid at
$r=\infty$.
We extract the radiated energy-momentum at finite ($r=50M, 60M,
\cdots, 100M$)
radius and extrapolate to $r=\infty$ using both linear and quadratic
extrapolations. We use the difference of these two extrapolations as
a measure of the error.

\begin{table}[t]
  \caption{Final spin parameters for Bowen-York data (BY), the new data at
 $\Lambda^{\rm max}$, and conformal Kerr data. 
Note that $\alpha_H = S_H/M_{H\infty}^2$,
$\alpha_{\rm rad}= J_{\rm ADM}/(M_{\rm ADM}-\delta M_{\rm rad})^2$.
The  notation of the form  $1.23(4\pm5)$  used below is shorthand for
$1.234\pm0.005$.}
   \label{tab:remParamters}
\begin{ruledtabular}
\begin{tabular}{|l|l|l|l|}
ID	& $\alpha_H$ & $\alpha_{\rm rad}$ &  $10^3\delta M_{\rm
rad}$\\
\hline
BY                   & $0.9309(0\pm2)$ & $0.9308(4\pm1)$ &
$1.47(6\pm5)$
\\
$\Lambda^{\rm max}$ & $0.935(2\pm1)$ & $0.9351(3\pm1)$ & $1.49(1\pm6)$
\\
cKerr & $0.9348(2\pm1)$ & $0.9347(6\pm2)$ & $1.49(0\pm7)$
\\
\hline
\hline
ID	& $M_{H_\infty}$ & $S_H$ & $M_{\rm irr}$ \\
\hline
BY &    $1.0365(1\pm3)$ & $ 1.0001(5\pm5)$ & $ 0.8563(6\pm3)$\\
$\Lambda^{\rm max}$ & 
$ 1.0341(5\pm7)$ & $1.000(2\pm1)$ & $0.850(9\pm1)$\\
cKerr &  $1.0343(1\pm2)$ &  $1.0000(8\pm4)$ & $0.85138(9\pm5)$
\end{tabular}
\end{ruledtabular}
\end{table}

The initial 3-metric  for these black holes is equivalent
to the Kerr 3-metric  (in quasi-isotropic coordinate) with a
non-trivial axisymmetric distortion wave.
This initial distortion contains energy that is partially radiated
to infinity and partially absorbed by the black hole. From
Table~\ref{tab:IDparameters} we see that  in the extreme case
2.8\% of the total mass lies outside the black hole (here $S=1$)
\bea
\label{eq:wavemass}
M_{\rm ADM}-M_{H_0}
&=& \sqrt{\frac{S}{\epsilon_S}}-\sqrt{\frac{S}{\chi}}
\nonumber \\
&=& 1.0356-1.0073
\nonumber \\
&=& 0.0283 \,.
\eea
This difference, $E_{\rm wave} = M_{\rm ADM}-M_{H_0}$, is the energy 
associated with the distortion wave. From
Table~\ref{tab:remParamters}, we see that 95\% of the energy
in the wave is eventually  absorbed by the black hole. Of the total ADM mass,
only 0.15\% is actually radiated to infinity. 
This means that the black hole transitions from a nearly maximally
spinning object to a submaximal Kerr black hole with spin 0.935,
as shown in the above Table~\ref{tab:remParamters}.
This effect is also shown in Fig.~\ref{fig:AH},
where, at around $t=10.7M$ of the evolution, a new external AH
appears suddenly  with a larger area and much larger coordinate radius.
Initially, only one trapped surface exists, with a very small
coordinate
radius of  $0.0095M$.
As the evolution proceeds this horizon acquires
an oblate form with the equatorial radius nearly double the polar
radius. The jagged features associated with the smaller trapped
surface at later times are due to poor resolution of this very small
object (which reduces in polar radius from $\sim 0.009M$ to $<0.002M$). 
At $t=10.7M$ an outer AH forms with polar radius
$\sim0.14M$ and equatorial radius $0.27M$. 
At the same time a third, shrinking
trapped surface also appears that eventually meets the original smaller
trapped surface, after $t=28.2M$. These newly formed
horizons are oblate. 
We can also observe notable quasinormal oscillations 
soon after their formation.

In order to understand the jump in the area in Fig.~\ref{fig:AH} when
the initial distortion is absorbed by the rotating hole
we note that
\bea
\frac{A_H}{S} &=& \frac{8\pi}{\xi}
\nonumber \\ 
&=& 8\pi\,\frac{1+\sqrt{1-\chi^2}}{\chi}\,,
\eea
where $A_H$ is the area of the horizon.
By keeping $S$ fixed and allowing $M_{H}$ to vary, we obtain
\bea
\frac{\delta A_H}{A_H}=
\frac{2}{\sqrt{1-\chi^2}}\left(\frac{\delta M_H}{M_H}\right) \,.
\eea
In our case this leads to a relative increase in the horizon area of
$15\%=5.64\times 2.685\%$
(using the numbers of Table~\ref{tab:remParamters}) 
in agreement with the $12.4-18.1\%$ jump observed
in the bottom plot of Fig.~\ref{fig:AH} (from 29.8 initially to 34 and then
to 36.4) for the $A_H$.
Another interesting aspect of these horizons is the jump in the
deviation from nearly spherical to spheroidal. We define the deviation
functions $D$ to  measure the non-spherical shape of the horizons.
We define the coordinate dependent deviation  $D_r=(r_{\rm equatorial}-r_{\rm
polar})/(r_{\rm equatorial}+r_{\rm polar})$, where $r$ is the
coordinate radius, and coordinate independent deviation
$D_c=(c_{\rm equatorial}-c_{\rm polar})/(c_{\rm equatorial}+c_{\rm
polar})$, where $c$ is the invariant measure of the circumference
(see~\cite{Alcubierre:2004hr} for formulas relating $D_c$ to the intrinsic spin $a/M$
for Kerr).
For a spherical horizon, $D_c$ must be zero in all coordinate systems,
while $D_r$ can differ from zero, with the difference being a measure
of how distorted the coordinate system is in the neighborhood of the
AH.

\begin{figure}[!ht]
  \includegraphics[width=\columnwidth]{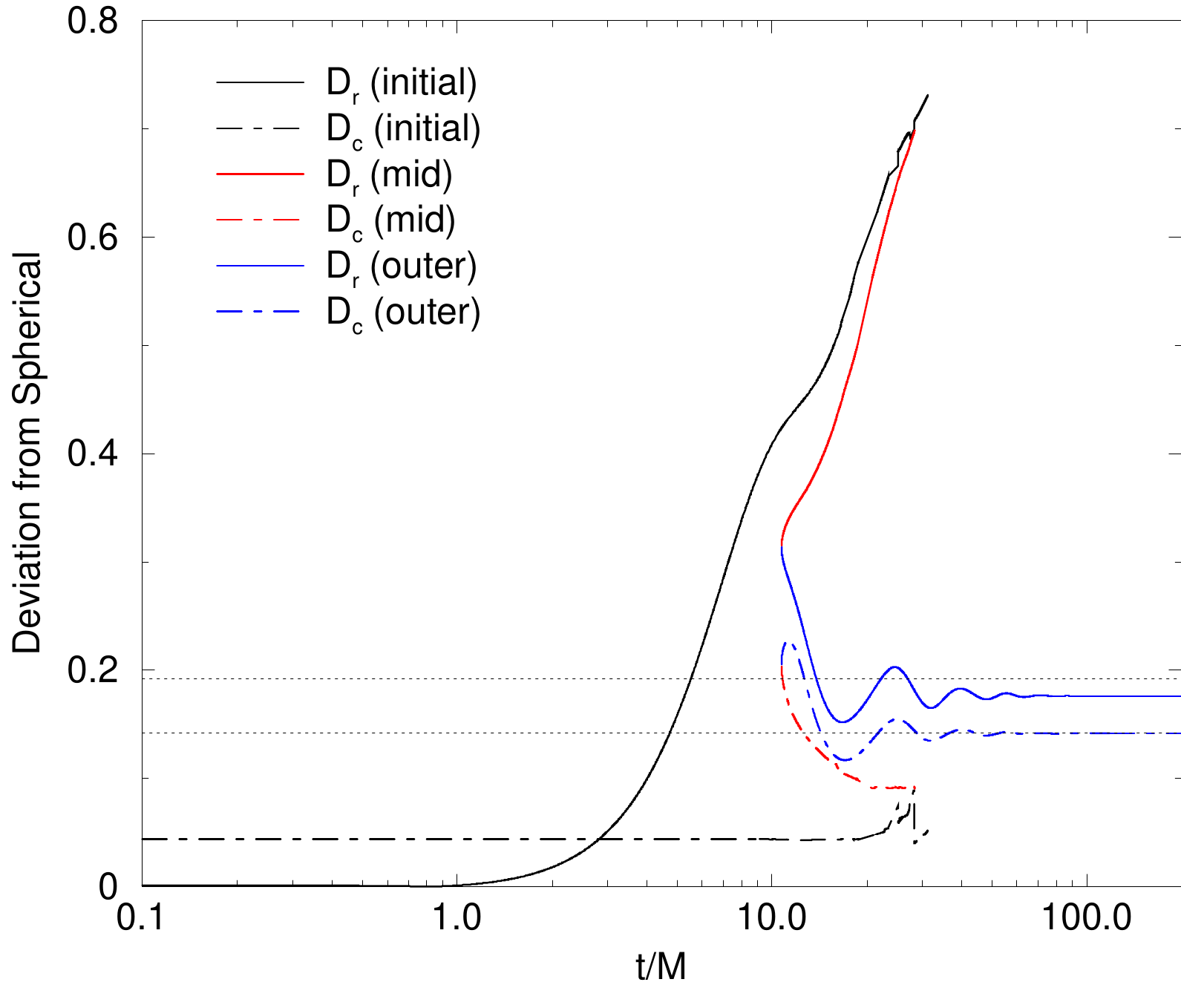}
  \caption{The deviation of the AH shape from a pure sphere $D$, for
the 
$\Lambda=0.7831$ data, as
measured using the coordinate radii $D_r = (r_{\rm equatorial} -
r_{\rm polar})/(r_{\rm equatorial} + r_{\rm polar})$ and the
coordinate independent measure using the equatorial and polar
circumferences $D_c = (c_{\rm equatorial} -
c_{\rm polar})/(c_{\rm equatorial} + c_{\rm polar})$. Note how the
initial horizon starts out as a coordinate sphere, which rapidly
becomes oblate. While the invariant circumferences indicate a nearly
constant deviation from spherical even at t=0. The dotted lines show
the Kerr values for $D_c$ for the initial intrinsic spin of
$\chi_0=0.9856$ (upper) and the equilibrium spin of $\chi_\infty=0.9352$
(lower).
}
\label{fig:AHshape}
\end{figure}

The initial horizon is
highly spinning. Therefore, we would expect it to show the largest
deviation from spherical. However, as seen in Fig.~\ref{fig:AHshape},
this horizon is more spherical (i.e.\ $|D_c|$ is smaller)
 than the slower spinning equilibrium
horizon (slower in the sense $S/M_H^2$ is smaller, $S$ itself is
constant). Also note that the distortion of the coordinates, as
measured by the relative difference between the $D_c$ and $D_r$ are much smaller
for the equilibrium horizon despite the fact that this horizon is less
spherical. 
We also observe a notable constancy of $D_c$ of
the smallest horizon, beginning from the initial time slice, through the formation of
the outer apparent horizon, even until later times when the 
shrinking internal horizon meets the inner horizon
at around $t=28.2M$. As expected the outer horizon
is  very distorted when it first forms  and then settles
 to the $D_c$ of a Kerr
horizon with the corresponding spinning parameter $a/M=0.935$.

\begin{figure}[!ht]
  \includegraphics[width=\columnwidth]{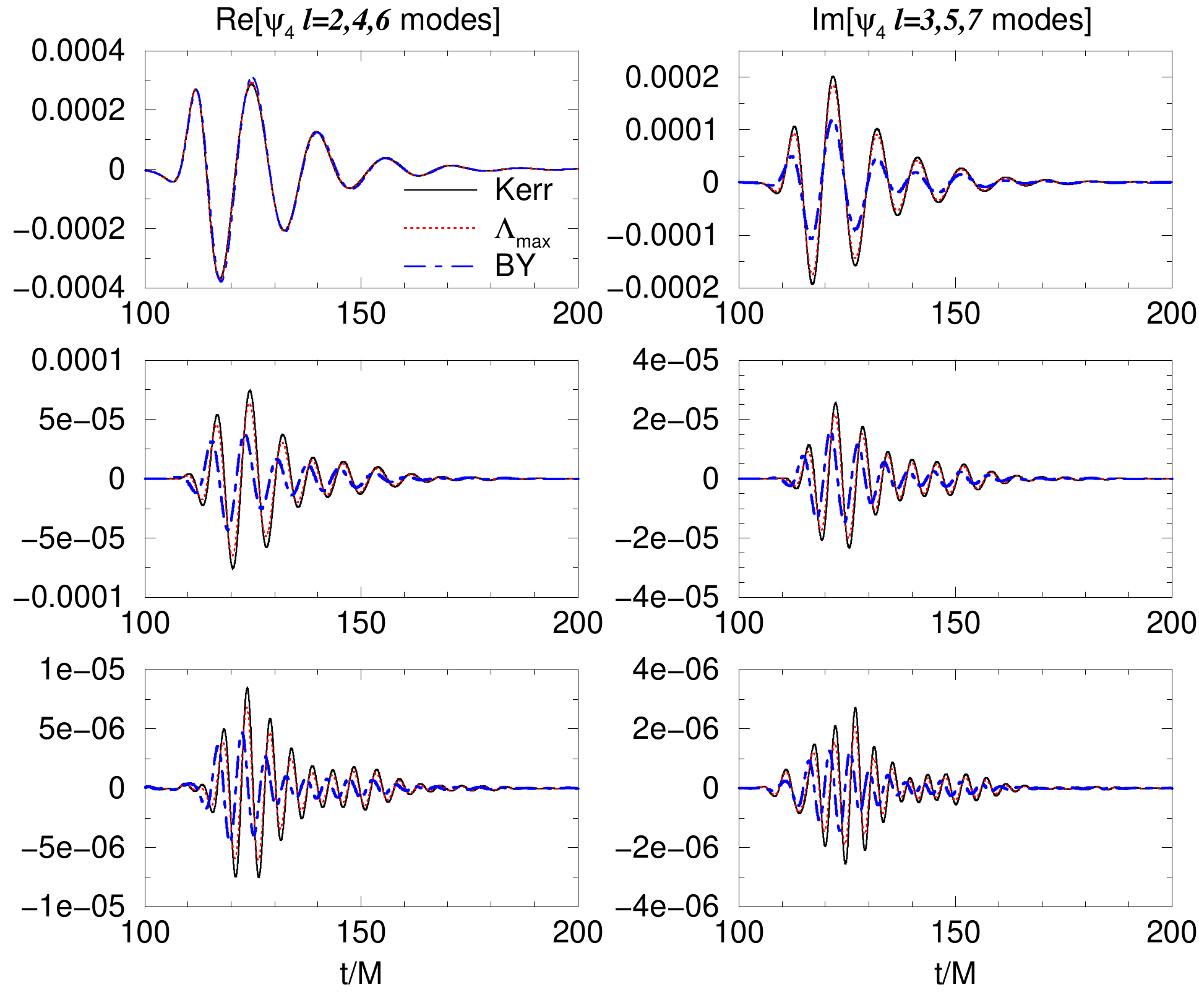}
  \caption{The $(\ell,0)$ modes of $\psi_4$ for the BY, $\Lambda^{\rm
max}$, and cKerr data. The plot is arranged in two columns with even
$\ell$ modes to the left and odd $\ell$ modes to the right. The plots
are arranged vertically with lower
$\ell$ modes above higher $\ell$ modes. For the even $\ell$ modes, the
odd part of $\psi_4$ is zero and not shown, while for odd $\ell$
modes, the even part of $\psi_4$ is zero and not shown. Note how the
different initial data give essentially the same $\ell=2$ signal, but
differ quantitatively in the higher $\ell$ modes.
}
\label{fig:wave}
\end{figure}

\begin{figure}[!ht]
  \includegraphics[width=\columnwidth]{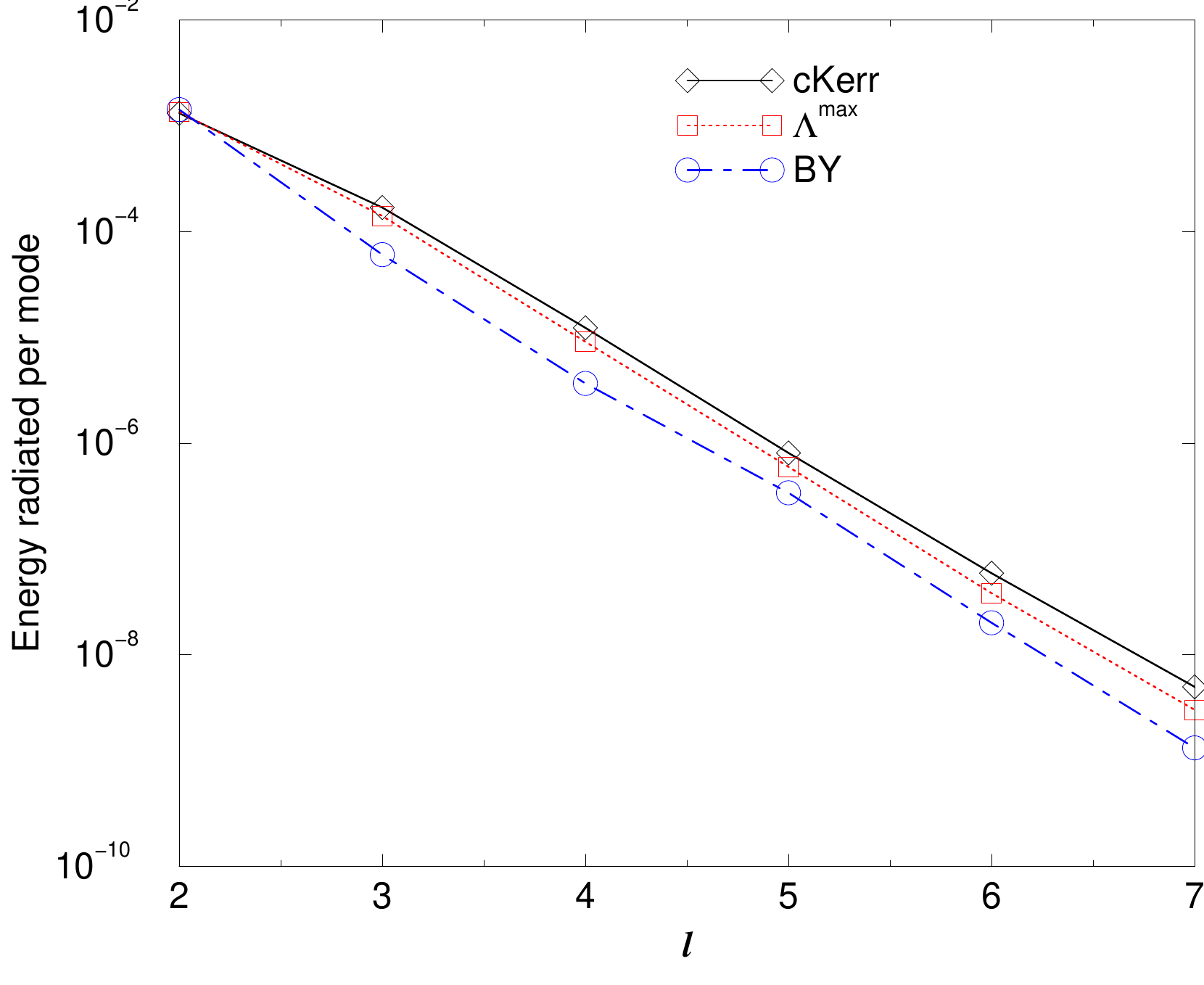}
  \caption{The energy per $(\ell,0)$ mode for the BY, $\Lambda^{\rm
max}$, and cKerr data.}
\label{fig:energy}
\end{figure}

In Figs.~\ref{fig:wave}~and~\ref{fig:energy} and
Table~\ref{tab:e_v_l}, we show the waveform and
radiated energy per $\ell$ mode for pure BY, $\Lambda^{\rm max}$ and
cKerr extrinsic curvature data. Note how for all modes with $\ell>2$, the cKerr data radiated the
most, while pure BY data radiates the least. The effect, however, is
essentially in the higher $\ell$ modes. For the $\Lambda^{\rm max}$ data, 
the ADM mass
is lower than for BY, and the system radiates more than BY. Both
effects lead to larger final spins for the  $\Lambda^{\rm max}$ data
when compared to BY data. The $\Lambda^{\rm max}$ data, however, 
does not radiate more than the cKerr data. Thus there is a trade-off
between the lower ADM mass of the $\Lambda^{\rm max}$ data and 
total mass radiated for the cKerr data. Slightly smaller final spin
could be obtained by optimizing the $\Lambda$ parameter, but the
effect will be quite small. Interestingly, the BY data radiates more
in the $\ell=2$ mode than the other data. It is only after summing
over all modes, that the other data sets end up radiated more.


\begin{widetext}

\begin{table}[t]
  \caption{The energy radiated ($\delta E/M$), per $\ell$ mode, for the BY,
$\Lambda^{\rm max}$, and cKerr data. Note that the BY data radiates
the most in the $\ell=2$ mode, but least in all other modes. The total
radiated energy ($\delta E/M$) over all modes is
$(1.4799\pm0.0010)\times 10^{-3}$, $(1.4978\pm0.0010)\times 10^{-3}$,
$(1.4983\pm0.0010)\times 10^{-3}$ for BY, $\Lambda^{\rm max}$, and
cKerr, respectively.}
   \label{tab:e_v_l}
\begin{ruledtabular}
\begin{tabular}{|l|l|l|l|}
Mode & BY &  $\Lambda^{\rm max}$ & cKerr \\
\hline
(2,0) & $(1.416\pm0.001)\times10^{-3}$ & $
(1.349\pm0.001)\times10^{-3}$ & $(1.318\pm0.001)\times 10^{-3}$\\
(3,0) & $(6.01\pm0.01)\times10^{-5}$ & $(1.392\pm0.005)\times10^{-4}$ &
$(1.672\pm0.007)\times10^{-4}$\\
(4,0) & $(3.56\pm0.11)\times 10^{-6}$ & $(8.85\pm2.8)\times 10^{-6}$ &
   $(1.200\pm0.039)\times 10^{-5}$ \\
(5,0) & $(3.12\pm0.30)\times 10^{-7}$ & $(5.38\pm0.60)\times 10^{-7}$&$
(7.25\pm0.86)\times10^{-7}$\\
(6,0) & $(1.13\pm0.89)\times 10^{-8}$ & $ (2.63\pm1.22)\times 10^{-8}$
& $ (3.88\pm2.03)\times10^{-8}$\\
(7,0) &  $(0.3\pm1.0)\times10^{-9}$ & $(1.19\pm1.82)\times 10^{-9}$ &
$(1.39\pm3.58)\times10^{-9}$

\end{tabular}
\end{ruledtabular}
\end{table}

\end{widetext}

\section{Discussion}

In this paper we studied conformally flat initial data
with an extrinsic curvature ansatz that interpolates between the
Bowen-York solution and the conformal Kerr solution.
We found that, given our ansatz, the maximum intrinsic spin, normalized
by the ADM mass,
does not correspond to a pure conformal Kerr extrinsic curvature,
but rather to a weighted sum of roughly 78\% conformal Kerr and 22\% Bowen-York extrinsic
curvatures. While if we normalize the intrinsic spin with the horizon
mass,
the maximum occurs at roughly  75\% conformal Kerr  and 25\% Bowen-York
extrinsic curvatures.
This result can be interpreted as implying that,
even if the conformal Kerr extrinsic curvature is the ideal choice for
conformally Kerr initial data (3-metric), for conformally flat initial data,
a mixture including Bowen-York compensates
in part for the distortions produced by conformal flatness, leading to
slightly larger intrinsic spins.

Other numerical explorations are possible using Eq.~(\ref{eq:Komega}) 
above for different choices of $\omega$.
One could also try a variational approach, where 
an extremum in the ADM mass can be found by varying the function $\omega$, but this is beyond the scope of the current paper.
Note that Dain and Friedrich~\cite{Dain:2001ry} have described the
general form of the extrinsic curvature for asymptotically flat initial data.

Another example of conformally flat initial data with different
extrinsic curvature is provided by the thin-sandwich approach
\cite{Pfeiffer:2005jf}, but this requires solving a system of five
coupled partial differential equations.  Our approach is rather to try
to keep  all the benefits of the simplicity of the Bowen-York
solutions, i.e., an analytic conformal extrinsic curvature,
Eqs.~(\ref{eq:Kaa})-(\ref{eq:lambda}), with a very simple form that is
easy to incorporate in any numerical code that solves the Hamiltonian 
constraint for the conformal factor.  It is the intentions of the
authors to deliver an open source code incorporating these new data to
the Einstein Toolkit collaboration~\cite{Loffler:2011ay}.

Ultimately we are interested in the evolution of these initial data and
the  ratio $\chi=S/M_H^2$ at late times, since this represents the
actual spin of the black holes in a black-hole-binary simulation (that
is, the individual black holes will equilibrate on much smaller
timescales than the binary inspiral). 
 Our numerical evolutions
find that only an small fraction of the available energy initially 
outside the black hole
radiates to infinity. Thus from the two initial data indicators,
$\epsilon_S$ and $\chi$ the former, which is normalized by the ADM mass
is the most accurate.
We thus conclude that our family of initial data is limited to represent
black holes with intrinsic spins $\alpha^{\rm max}<0.9352$.

We also find an important jump in the apparent horizon location in numerical 
coordinates when radiation falls in and the hole settles from an extreme
near-spherical horizon to a submaximal Kerr hole with a much larger coordinate
size (an increase in coordinate radius of over a factor of 30). 
One needs to take into account these jumps both in coordinate size and physical
mass when designing simulations
and especially when measuring properties such as
the mass, spin, linear
momentum~\cite{Krishnan:2007pu},
and when tracking the motion of binary horizons in a full numerical
simulation. 
More invariant measures of the horizon properties like its area,
Fig.~\ref{fig:AH}, and ratio of polar to equatorial circumferences, 
Fig.~\ref{fig:AHshape}, still display an initial
quasi-spherical (small distortion) AH, with a discontinuous jump
at  $t=10.7M$ to
a larger, highly-distorted horizon that, after some oscillation,
 settles to the spherical
deviations
corresponding to that of a Kerr hole with $a/M=0.935$.
The moving puncture approach also allows us to keep track of the internal
horizon connecting the initial and final ones even after the formation of
the latter at $t=10.7M$.
Although the details of this horizon transition are specific of the 
problem we studied, any other initial data set for multi black holes is 
expected to have a radiation content that is partially absorbed and partially
radiated to infinity, and hence present similar qualitative early evolutions
to those presented here.

\section{acknowledgments}

We thank S. Dain and O. Ortiz for valuable discussions on the subject
of this paper.
We gratefully acknowledge the NSF for financial support from Grants
No. PHY-0722315, No. PHY-0653303, No. PHY-0714388, No. PHY-0722703,
No. DMS-0820923, No. PHY-0929114, PHY-0969855, PHY-0903782,
and No. CDI-1028087; and NASA for financial support from NASA Grant 
No. 07-ATFP07-0158.  Computational resources were
provided by the Ranger cluster at TACC (Teragrid allocation TG-PHY060027N) 
and by NewHorizons at RIT.

\bibliographystyle{apsrev}
\bibliography{../../../Bibtex/references}

\end{document}